# Aplicabilidad de los Modelos de Madurez de Business Intelligence a PYMES

González-Varona JM, López-Paredes A, Pajares J, Acebes F, Villafáñez F

**Resumen**

Las grandes empresas están totalmente involucradas en su transformación digital, y en concreto en el desarrollo de proyectos estratégicos de Business Intelligence (BI). Cuentan con una Estrategia Digital y directivos al más alto nivel gestionando el cambio. En las PYMES también se realizan proyectos de BI. En este trabajo se presentan los resultados de un estudio interpretativo realizado sobre una muestra de PYMES de diferentes sectores, que están realizando proyectos de BI. Estudiamos si los modelos de madurez en BI son válidos para las PYMES y su utilidad para analizar los proyectos y evaluar a su finalización los resultados.

## 1. Introducción

Las nuevas tecnologías digitales suponen uno de los mayores retos a los que se enfrentan las empresas en la actualidad (Hess et al. 2016). Ninguna es inmune a sus efectos y de la integración y explotación de estas tecnologías dependerá su desempeño futuro. La integración en nuestra sociedad de las denominadas tecnologías SMAC: Social (Redes sociales), Mobile (Dispositivos móviles), Análisis de datos (Big Data) y operaciones en la nube (Cloud), está generando una digitalización progresiva y sin precedentes que fomenta la innovación y transformación de las empresas y sociedad (Legner et al. 2017).

La digitalización constituye una fuente de oportunidades, pero también una amenaza a la supervivencia de aquellas empresas que no sean capaces de adaptarse. Para Legner et al. (2017) la integración de las tecnologías SMAC complementa y/o enriquece los productos y servicios existentes y permite construir modelos de negocio completamente nuevos. Hess et al. (2016) indican que el potencial de cambio generado por las nuevas tecnologías digitales va más allá de productos, procesos, canales de venta o de suministro: los modelos de negocio se están rede iniendo e incluso cambiando por completo.

La Transformación Digital (TD) de las organizaciones supone cambios estructurales que generarán mejoras futuras en la competitividad y posición de las organizaciones, pero que supone asumir riesgos a corto plazo que pueden afectar a la rentabilidad (Sánchez-Marín et al. 2018). Es por ello que normalmente son las organizaciones de mayor tamaño las que tienen capacidad para asumir los riesgos derivados de la TD, mientras que las organizaciones pequeñas y medianas son más cautas.

En el Informe PYME España 2018 de FAEDPYME (Sánchez-Marín et al. 2018) se indica que las PYMES españolas tienen un nivel básico de digitalización. Además, existe una diferencia clara entre pequeñas y medianas empresas. Las medianas empresas suelen alcanzar el tamaño mínimo suficiente para incorporar habilitadores de la TD, como plataformas e-commerce, mientras que las pequeñas empresas aún no tienen la capacidad para avanzar en la TD. Aunque la TD es un hecho ineludible el informe señala que las PYMES españolas aún no son realmente conscientes de su importancia y por ello no incluyen todavía la TD como una prioridad. Por último, en cuanto a la estrategia de digitalización, las PYMES sí se dan cuenta de que la TD afecta a toda la estructura de la empresa, a sus clientes, proveedores, trabajadores, etc. y que por ello la alta dirección de la organización tendrá que comprometerse con los procesos de cambio que de forma inevitable afectarán a la organización.

En el informe de Análisis dinámico del tejido empresarial de Castilla y León (CES, 2017) [4] se pone de relieve que las PYMES, y sobre todo microempresas y pequeñas empre-

✉ González-Varona JM * y **
josemanuel.gonzalez.varona@uva.es

López-Paredes A ** y ***

Pajares J ** y ***

Acebes F ** y ***

Villafáñez F ** y ***

* Director de Proyectos, CGPROTEC S.L. (www.cgprotec.com), Edificio CT  A, Campus Miguel Delibes, 47011, Valladolid.

** Dpto. de Organización de Empresas y CIM. Escuela de Ingenierías Industriales. Universidad de Valladolid. C/ Paseo del Cauce 59, 47011 Valladolid (Spain).

*** INSISOC, Edificio U  AInnova, Paseo de Belén 9. 47011 Valladolid. www.insisoc.org

[4] CES, I. C. E. y S. de C. y L. (2017). Análisis dinámico del tejido empresarial de Castilla y León. Descripción del panorama actual, factores determinantes y líneas de actuación.

sas, forman el tejido empresarial relevante de la Comunidad Autónoma. El 53,3% de las empresas no tiene empleados, el 42,2% son microempresas con entre 2 y 3 empleados, el 4% son pequeñas empresas, que tienen una media de 22 trabajadores, y las medianas y grandes representan entre el 0,46% y el 0,09% de media, teniendo por tanto un peso poco relevante dentro del tejido empresarial de la Comunidad Autónoma; aunque son estas empresas medianas y grandes las que tienen una mayor permanencia en el tiempo.

En la actualidad las PYMES se enfrentan a problemas derivados de un volumen excesivamente grande de datos, y por contra falta de información y falta de conocimiento para tratar la complejidad que afecta a la empresa. En consecuencia, los directivos de las PYMES utilizan principalmente su experiencia para tomar decisiones, lo que implica un alto riesgo de fracaso (Papachristodoulou et al. 2017). Las herramientas de Business Intelligence (BI) dan forma a una nueva generación de Sistemas de Soporte a la Decisión (SSD) que han demostrado ser muy útiles para la toma de decisiones generando ventajas y beneficios a las empresas.

En los últimos años, la TD se ha convertido en un fenómeno importante en la investigación de los sistemas de información (Bharadwaj et al. 2013; Piccinini et al. 2015), así como para todos aquellos profesionales de empresas preocupados por la competencia creciente en los mercados (Fitzgerald et al. 2014; Westerman et al. 2011).

La implantación de proyectos de BI se convierte en un factor estratégico en tanto que tiene capacidad para generar ventajas competitivas: permite obtener información para responder a los retos de entrada en nuevos mercados, planificación de la producción, eliminación de duplicidades de gestión, evitar islas de información, rentabilidad, etc. Más aún, de acuerdo con Stone and Woodcock (2014) los consumidores son cada vez más digitales, compran, interactúan digitalmente, etc. y las organizaciones no pueden renunciar a su digitalización. En general esto implica una transformación completa, que difícilmente puede abordarse de forma radical. Los modelos de madurez permiten a las empresas ir evolucionando de forma progresiva en aquellas capacidades que les permitan afrontar con éxito la TD (Lorenzo Ochoa 2016).

En este trabajo hemos realizado el análisis y la evaluación de la implantación de varios proyectos de Business Intelligence en PYMES de Castilla y León. Hay tres cuestiones que constituyen el objeto de la investigación:

RQ1 - ¿son los modelos de madurez en BI definidos por consultoras apropiados para aplicarlos a PYMES?

RQ2 - ¿son de utilidad estos modelos para analizar y seleccionar los proyectos de BI más adecuados y para evaluar su impacto al concluir su ejecución?

RQ3 - ¿en qué medida los proyectos de BI contribuyen a alcanzar la madurez digital, concepto propuesto por Kane (2017) para referirse a la capacidad de responder al cambio de manera temprana?

El resto del documento se ha organizado de la siguiente manera: en la sección 2, abordamos el marco teórico de los sistemas de BI. En la sección 3, describimos el método de investigación, detallando el proceso de recopilación y análisis de datos. En la sección 4 se exponen los resultados del estudio interpretativo desarrollado en las empresas seleccionadas, y finaliza os presentando las principales conclusiones en relación con las tres cuestiones objeto de este trabajo de investigación

## 2. Marco teórico: Business Intelligence (BI)

A pesar de haber cobrado gran importancia en las últimas décadas, el concepto Business Intelligence no es nuevo, aunque sí que ha evolucionado a gran velocidad. El término fue introducido por Hans Peter Luhn (P. Luhn 1958) para referirse a la habilidad de aprender de las interrelaciones de los hechos identificados de tal forma que guíen nuestras acciones hacia el objetivo deseado. Howard Dresner en 1989, antiguo analista y vicepresidente del grupo Gartner, definió BI como un término que describe los conceptos y métodos para mejorar la toma de decisiones en el negocio mediante el uso de sistemas de soporte a la decisión basados en hechos.

Para Reinschmidt and Francoise (2000) un sistema de BI es "un conjunto integrado de herramientas, tecnologías y productos programados que se utilizan para recopilar, integrar, analizar y hacer que los datos estén disponibles". Zeng et al. (2006) definen BI como “el proceso de recogida, tratamiento y difusión de información que tiene como objetivo la reducción de la incertidumbre en la toma de decisiones estratégicas”.

Existe coincidencia en incidir en la importancia de la información para la toma de decisiones en la empresa, y relacionar el BI con las tecnologías y aplicaciones que se utilizan para recoger y analizar datos e información. Tvrdíková (2007) describe como característica básica de BI la ca-pacidad para recopilar datos de fuentes heterogéneas, poseer métodos analíticos avanzados y poder soportar demandas de múltiples usuarios.

Dedić and Stanier (2016) amplian el concepto de BI de su dimensión instrumental a la funcional, y consideran que abarca las “estrategias, procesos, aplicaciones, datos, productos, tecnologías y arquitecturas técnicas empleadas para sustentar la recolecta, análisis, presentación y difusión de la información del negocio”.

Gartner[5] define BI como el conjunto de "aplicaciones, infraestructuras y herramientas y las buenas prácticas que permiten acceder y analizar la información para mejorar y optimizar tanto las decisiones como el desempeño" y Forrester[6] de forma similar se refiere al "conjunto de metodologías, procesos, arquitecturas y tecnologías que mejoran las salidas de los procesos de gestión de información para el análisis, realización de informes, gestión del desempeño y distribución de la información".

Una vez analizadas estas definiciones observamos coincidencia en la visión actual de los sistemas de BI:

- el objetivo fundamental es mejorar la toma de decisiones dentro de la empresa,
- engloban todas las herramientas que le permitan analizar, estructurar o procesar información.

Si bien la importancia de los sistemas de BI para proporcionar información relevante para la toma de las mejores decisiones, especialmente en mercados con una alta competencia, ya ha sido señalada (Popovič et al. 2012), aún no han sido suficiente ente investigados los beneficios que aporta el BI a las empresas, especialmente los beneficios a medio y largo plazo.

Según Zeng et al. (2006) una empresa aplicará exitosamente un sistema de BI si utiliza correctamente datos válidos, integrados, y en tiempo, además de las herramientas que le permiten transformar los datos en información relevante para la toma de decisiones. Para Glazer (1993) las empresas exitosas se centran en capturar el valor de la información a lo largo de toda la cadena de valor de la información y no en la tecnología (que sí tiene incidencia en cuanto a incrementar la velocidad y forma de transferir la información o la cantidad de información que el sistema puede procesar).

Aunque el BI agrega valor principalmente al comienzo de la cadena de valor de la información (Popovič et al. 2012) los beneficios derivados de la implantación de un proyecto de BI se manifiestan, normalmente a largo plazo. El hecho de ser indirectos y muy distantes en el tiempo hace complicado evaluar el resultado efectivo.

Las probabilidades de éxito en la implementación de un proyecto de BI se multiplican cuando se identifican desde un principio las necesidades de negocio específicas de la organización y además esas necesidades se usan para orientar la naturaleza y alcance de los proyectos. Yeoh and Koronios (2010) definen un marco para la implementación del BI en el que identifican un conjunto de factores críticos de éxito. Del análisis de los resultados obtenidos se deduce que los factores no técnicos, como los organizacionales y los relacionados con los procesos, son más influyentes e importantes en el éxito de la implementación que los factores de tipo tecnológico y relacionados con los datos. La conclusión más relevante es que la implementación de los proyectos de BI se debe orientar a los negocios y estar centrada en la organización.

Los factores organizacionales son los que contribuyen en mayor medida a modificar la opinión y la motivación interna de los usuarios para la aceptación y utilización del BI. Si los usuarios perciben beneficios tangibles, particularmente con respecto a su imagen y que los resultados de uso son visibles y demostrables, comenzarán a usar BI intensivamente (Grublješič and Jaklič 2015a). Si perciben mejoras en su imagen personal que potencialmente supongan una mayor inclusión social y estatus en la organización, será más probable que acepten y comiencen a usar el BI. La influencia social y la demostrabilidad de los resultados tienen reconocida su relevancia en el proceso cognitivo de un individuo y en su comportamiento (Venkatesh et al. 2003; Venkatesh and Bala 2008). Esta motivación influye especialmente en la aceptación de uso del BI si tenemos en cuenta que normalmente su utilización es voluntaria.

Popovič et al. (2012) desarrollan un modelo que contribuye a explicar las relaciones entre los determinantes del éxito de un sistema de BI, en especial la influencia de la madurez de BI y el establecimiento de una cultura de toma de decisiones basadas en datos. Inicialmente las organizaciones han de afrontar dos cuestiones fundamentalmente para desarrollar su proyecto de BI. Por un lado, habrá gran cantidad de datos de fuentes heterogéneas que es necesario analizar e integrar (Elbashir et al. 2008). Por otro lado han de proveer de capacidades de análisis de datos para el análisis de datos de negocio (Trkman et al. 2010). Estas dos dimensiones del BI inciden en el desarrollo de la madurez del BI.

Un incremento de la madurez del BI propiciará la obtención de una información de mayor calidad (Hannula and Pirttimaki 2003); en dos aspectos relevantes, tanto en el contenido de la información como en un acceso de calidad a la información, un incremento de la percepción de calidad de la información por parte de los usuarios se traducirá en un mayor uso del BI en los procesos de negocios. Además el desarrollo de una cultura empresarial de toma de decisiones basadas en datos también afectará positivamente al uso de la información en los procesos de negocios.

Tener un presupuesto apropiado también es un determinante del éxito de la implementación de un proyecto de BI. Yeoh and Koronios (2010) indican que el éxito de un proyecto está determinado, en un 90%, antes del primer día, y dependerá de tener un alcance muy claro y bien comunicado, expectativas y plazos realistas, y un presupuesto apropiado.

[5] https://www.gartner.com/it-glossary/business-intelligence-bi/

[6] https://www.forrester.com/Business-Intelligence

# 3. Metodología de investigación

Para responder a los interrogantes planteados hemos llevado a cabo un estudio interpretativo de cinco casos de implementación de proyectos de BI en PYMES de Castilla y León. El trabajo se ha realizado entre enero y julio de 2019. Walsham (1993) afirma que los métodos de investigación interpretativa están "destinados a producir una comprensión del contexto del sistema de información, y el proceso por el cual los sistemas de información influyen y son influenciados por el contexto". Este método de investigación ya ha sido utilizado en diferentes estudios de investigación relacionados con BI (Yeoh and Koronios 2010; Li et al. 2013; Grublješič and Jaklič 2015b; Ponelis 2015).

Como punto de partida en los estudios interpretativos de casos se permite la utilización de teorías ya existentes, aunque según Walsham (1995) se requiere un grado considerable de apertura a los datos de campo y disposición a modificar los supuestos y teorías iniciales. Tal y como sugieren Klein and Myers (1999) seguiremos los siete principios para la investigación interpretativa de campo (ver tabla 1).

**Tabla 1** Adaptación de (Klein and Myers 1999), aplicación de los siete principios de la investigación interpretativa a los casos de estudio.

| Principios | Aplicación de los principios a los casos de estudio |
|---|---|
| **1. El Principio Fundamental del Círculo Hermenéutico:** Este principio sugiere que toda comprensión humana se logra mediante la iteración entre la consideración del significado interdependiente de las partes y el todo que éstas forman. Este principio de comprensión humana es fundamental para todos los demás principios. | Para la recopilación de los datos se realizaron entrevistas sucesivas a nuevos intervinientes y se tomaron datos de archivo. Además en el análisis de datos comparamos continuamente concepciones teóricas de la literatura de estrategias de TD y las conclusiones del análisis narrativo inicial, especialmente con respecto a los cinco niveles de madurez del modelo de madurez de BI de Gartner. |
| **2. El principio de contextualización:** Requiere una reflexión critica de los antecedentes sociales e históricos del entorno de la investigación, de modo que el público al que se dirige pueda conocer cómo surgió la situación actual que se está investigando. | Recopilamos información de documentos internos y datos públicos para establecer el contexto organizacional. Esto nos ayudó a rastrear la linea de tiempo general y los eventos clave de la evaluación e implementación de los proyectos de BI. Para entender el contexto hay que tener en cuenta las caracteristicas particulares de las PYMES (bajo presupuesto, importancia en la toma de decisiones del CEO, etc.) y su ubicación en Castilla y León. |
| **3. El principio de interacción entre los investigadores y los sujetos:** Requiere una reflexión critica sobre cómo se construyeron socialmente los materiales de investigación (o "datos") a través de la interacción entre los investigadores y los participantes. | En el estudio hemos descrito la forma recopilación y análisis de los datos obtenidos. El enfoque estuvo consensuado entre el investigador principal, de campo, y un segundo investigador. |
| **4. El Principio de Abstracción y Generalización:** Requiere relacionar los detalles idiográficos revelados por la interpretación de los datos a través de la aplicación de los Principios 1 y 2 con conceptos teóricos generales que describen la naturaleza de la comprensión humana y la acción social. | El análisis de los datos obtenidos comprende la generalización por comparación con el modelo de madurez de BI de la consultora Gartner, que permite interpretar datos empiricos obtenidos en el principio de interacción. |
| **5. El principio de razonamiento dialógico:** Requiere sensibilidad a las posibles contradicciones entre las ideas preconcebidas teóricas que guian el diseño de la investigación y los hallazgos reales ("la historia que cuentan los datos") con ciclos posteriores de revisión. | Hemos reflejado de forma critica los antecedentes de nuestro estudio basados en la literatura existente de Business Intelligence, estrategia de transformación digital y sistemas de información. Para conseguir una mejor compresión de nuestros hallazgos empiricos hemos realizado varios ciclos de revisión que han hecho evolucionar nuestras conclusiones. |
| **6. El Principio de las Múltiples Interpretaciones:** Requiere sensibilidad a las posibles diferencias de interpretación entre los participantes, tal como se expresan tipicamente en múltiples narrativas o historias de la misma secuencia de eventos bajo estudio. Similar a los relatos de múltiples testigos, incluso si todos lo cuentan tal cual lo vieron. | En las entrevistas realizadas se incluyeron diversos perfiles no solamente directivos involucrados, tales como el CEO, Responables Informáticos, Marketing, etc. sino también empleados involucrados en la implementación. |
| **7. El principio de sospecha:** Requiere sensibilidad a posibles "sesgos" y "distorsiones" sistemáticas en las narraciones recogidas de los participantes. | Con el fin de evitar en la medida de lo posible los "sesgos" y "distorsiones" que pueden producirse en las narraciones cotejamos continuamente las narrativas de los participantes entre si, y sus interpretaciones con la información existente en nuestra base de datos. |

Como estrategia investigadora para recopilar datos en el estudio se optó por un enfoque cualitativo. Este tipo de estrategia se adopta cuando es necesario explorar un problema o área específicos, con frecuencia cuando se estudia a un grupo o población determinados. Creswell (2013) subraya su utilidad cuando necesitamos obtener datos mediante una interacción directa con la población objeto de estudio, a través de entrevistas o encuestas, para obtener un conocimiento detallado y completo.

Nuestro enfoque cualitativo se centró en el análisis de la madurez digital de las organizaciones objeto de estudio previo al inicio de la implantación de los proyectos de BI. El análisis previo incluyó la recopilación de documentación relevante para el proyecto, para su estudio y análisis, así como reuniones de toma de datos con interesados relevantes en el proyecto a través de conversaciones informales no estructuradas.

Según Myers and Newman (2007) las conversaciones informales no estructuradas, entrevistas cualitativas, suponen una excelente fuente de datos en investigaciones cualitativas de sistemas de información. La información ha sido suministrada por involucrados a distinto nivel en la implantación y uso posterior de un proyecto de BI. En la muestra utilizada no se incluyen organizaciones de cinco o menos miembros.

## 3.1. Recopilación de datos

El estudio de investigación comenzó con un análisis de la situación inicial de las PYMES respecto a su madurez digital y finalizó con el análisis de la situación de las PYMES con los proyectos ya implementados y la extracción de unas conclusiones finales. En todos los casos se ha podido incluir la información suministrada por los consultores que participaron en la implementación de los proyectos.

La obtención de los datos fue un proceso iterativo durante el tiempo que duró la implementación de los proyectos de BI. Una fuente importante de datos fueron conversaciones informales no estructuradas que no fueron grabadas por expreso deseo de los interlocutores, pero sí se documentaron mediante su transcripción inmediatamente después de la interacción. Se trata de una fuente importante de información según Schultze (2000) y nos permite superar las reticencias de muchos interesados relevantes y de los que de otra forma no podríamos recopilar información para la investigación.

**Tabla 2** Lista de entrevistados

| Cargo | Número de interacciones personales | | | | |
|---|---|---|---|---|---|
| | Empresa A | Empresa B | Empresa C | Empresa D | Empresa E |
| CEO / Gerente | 2 | 4 | 3 | 2 | |
| Coordinador técnico | | | 4 | | |
| Director Proyectos I+D+i | 5 | | | | |
| Resp. Producción | 3 | 4 | | | |
| Resp. Administración | 1 | 4 | 2 | | 2 |
| Resp. Marketing y diseño | | 1 | | | |
| Director comercial | | | | | 3 |
| Recursos humanos | 1 | | | | |
| Director de Calidad e innovación | | | | 3 | |
| Empleado. Mantenimiento | | | | | 3 |
| Empleado. Gestor de proyectos | | 1 | | | |
| Encargado. Toma de datos Centro A | | | 2 | 2 | |
| Encargado. Toma de datos Centro B | | | 2 | 2 | |
| Total de interacciones | 12 | 14 | 13 | 9 | 8 |

En línea con el principio fundamental del círculo hermenéutico para la investigación de campo interpretativa (Klein and Myers 1999) se realizaron entrevistas sucesivas a nuevos intervinientes durante todo el período de investigación, y se tomaron datos de archivo en base al interés de los investigadores por obtener un mayor número de perspectivas de los casos de estudio. Se entrevistó en total a 22 personas (ver tabla 2). Todas ellas estuvieron involucradas en la implantación de los proyectos de BI, desde los CEO y responsables de alto nivel, hasta trabajadores en las organizaciones objeto de estudio, además de los consultores que participaron en las implantaciones, que participaron en la validación de la información obtenida de las entrevistas. Algunos fueron entrevistados varias veces, de forma sucesiva, durante el proceso de investigación. Además, en el análisis de datos comparamos continuamente concepciones teóricas de la literatura de estrategias de TD y las conclusiones del análisis narrativo inicial, especialmente con respecto a los cinco niveles de madurez del modelo de madurez de BI de Gartner.

Las conversaciones se realizaron principalmente de forma presencial, cara a cara, en reuniones programadas con los interlocutores relevantes de las organizaciones objeto de estudio. Unas pocas se realizaron vía telefónica y no estaban programadas. La duración total de las interacciones, calculada como las horas registradas de reuniones presenciales y telefónicas fue de 84 horas, con 55 horas de entrevistas presenciales programadas y 29 horas de conversaciones telefónicas.

Para obtener datos de campo adicionales se realizaron visitas a las instalaciones de las PYMES. En total 9 visitas, que supusieron 12 horas de trabajo de campo: reuniones de equipo de implantación de BI, sesiones de formación en BI, y otras.

Además, en línea con el principio de contextualización (Klein and Myers 1999) también se procedió a la recolección de diversos documentos de datos para su estudio y análisis, tanto en formato papel como digital. Estos documentos fueron tanto de carácter interno facilitados voluntariamente por interesados en los proyectos de BI, como de carácter externo obtenidos a través de una revisión de diversas fuentes, tales como página web, noticias en periódicos, cuentas anuales públicas o presentaciones de empresa.

Se recopilaron en total 276 documentos, de ellos 110 documentos de carácter interno, que incluyeron presentaciones y comunicaciones internas relacionadas con el contenido, proceso y contexto del BI. Por otro lado, se recopilaron 166 documentos disponibles públicamente, que incluyeron informes anuales, presentaciones de la compañía, comunicaciones de prensa, etc. Los datos de archivo internos cubren todas las fases relevantes de la formulación e implementación de los proyectos de BI entre enero y junio de 2019. Los datos incluyen los relativos al año anterior al de la implantación del proyecto BI.

### 3.2. Análisis de los datos

El proceso de análisis de datos no fue lineal y emergente (Walsham 1995). Se incluyó un análisis narrativo de los casos de estudio (Myers 2013) que documenta cómo se desarrollaron en el tiempo. Estructurar la implantación de los proyectos de BI como una narración de forma cronológica nos muestra cómo unos eventos llevan a otros o crean las condiciones para otros. Utilizamos el marco de contexto proceso-contenido para la investigación transformacional propuesta por Pettigrew (1987, 2012) como guía para el análisis narrativo y como herramienta de análisis.

Para el análisis inicial utilizamos los datos de archivo recopilados, para más tarde depurar la narrativa incorporando nuevas ideas obtenidas de observaciones e interacciones personales. El análisis narrativo fue revisado por los CEO y/o responsables de implantación de los proyectos de BI, los cuales estuvieron de acuerdo con el contenido de los análisis. Al mismo tiempo se procedió a codificar la gran cantidad de datos cualitativos obtenidos, tales como notas de campo, transcripciones, etc. por uno de los investigadores de campo, de acuerdo con un segundo investigador independiente. Todas las discrepancias entre ambos se discutieron y resolvieron de forma consensuada.

Tal y como indican Kane et al. (2015) la TD es un fenómeno relativamente nuevo y aún no está claramente definida la situación final a la cual llegaría una organización digitalmente madura. Es por ello que es importante determinar cuál es la situación concreta en la cual se encuentran las organizaciones objeto de estudio respecto a su madurez digital. Las expectativas que genera la implantación de un proyecto de BI para una empresa dependerá de su estado de madurez (Caserio and Trucco 2018). La evaluación del estado de madurez de un sistema de BI podemos considerarlo como una medida de su calidad.

El análisis de los datos obtenidos comprende la generalización por comparación con el modelo de madurez de BI de la consultora Gartner, que permite interpretar datos empíricos obtenidos en el principio de interacción. Una vez obtenida la información relevante de cada una de las organizaciones en cuanto a sus avances en la digitalización de su actividad, procedimos a situar en uno de los niveles de madurez identi icados por la Consultora Gartner en su modelo de madurez de BI, a cada una de las organizaciones objeto de estudio.

Uno de los modelos más utilizados para realizar la evaluación del nivel de madurez de BI en las organizaciones es el facilitado por la consultora Gartner (2017), que realiza una actualización periódica. En su web, la empresa señala en diciembre de 2018 que hasta el 87% de las empresas cuentan con un nivel bajo de madurez digital[7]. La evaluación incluye tres áreas clave: personas, procesos, métricas y tecnología. Este modelo muestra 5 niveles de madurez (ver Figura 1):

1. Básico: Es el nivel más bajo. Existen datos desestructurados y mal interpretados. Los datos se procesan a través de hojas de cálculo y los informes son limitados. Faltan herramientas para medir el rendimiento y falta conciencia de la importancia del BI. Toda la gestión de la información y la elaboración de informes la realiza el departamento de TIC.

**Figura 1** Modelo de Madurez de Gartner.
*Elaboración propia.*

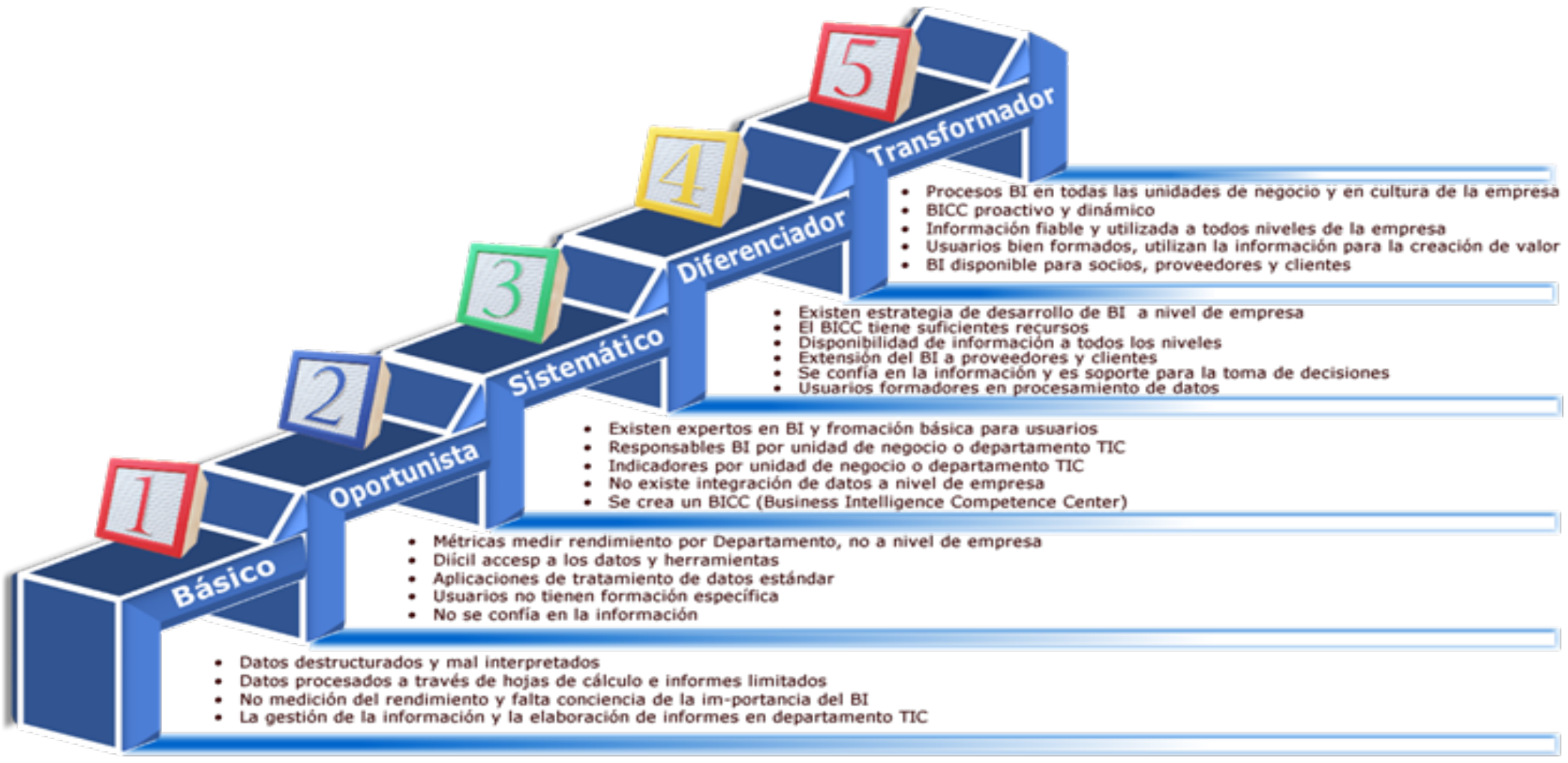

2. Oportunista: Utilización de métricas para medir el rendimiento a nivel departamental, pero no a nivel de empresa. Difícil acceso a datos y herramientas para su gestión. Las aplicaciones son estándar y los usuarios no tienen formación específica. Los responsables desconfían de la información resultante.

3. Sistemático: Ya existe personal en la empresa expertos en BI y tiene formación básica para usuarios. Responsables de BI por unidad de negocio o un responsable del departamento TIC. No existen indicadores para la empresa en su conjunto, aunque sí por unidad de negocio o departamento TIC. No existe una integración de los datos a nivel de empresa. Se crea el BI Competence Center (BICC).

4. Diferenciador: Existe una estrategia de desarrollo del BI en la empresa. El BICC tiene suficientes recursos para alcanzar sus objetivos. Disponibilidad de la información a todos los niveles y extensión de los procesos de BI a proveedores y clientes. Se confía en la información y se toma como soporte para la toma de decisiones. Los usuarios están formados en el procesamiento de datos y su utilización para la toma de decisiones estratégicas.

5. Transformador: Procesos de BI en todas las unidades de negocio y en la cultura de la empresa. El BICC es proactivo y dinámico. La información es fiable y se emplea a todos los niveles de la empresa. Usuarios bien formados y con acceso a la información, que utilizan para la creación de valor. El uso del BI está disponible para socios, proveedores y clientes.

En la etapa final del proceso de análisis de datos, creamos la Tabla 3: Madurez del BI de las empresas de la muestra seleccionada. En la Tabla 3 se representa la situación previa a los proyectos realizados, y la situación final alcanzada, tras concluir el estudio de los casos seleccionados.

## 4. Desarrollo de las entrevistas

Las cinco organizaciones seleccionadas para este trabajo fueron cuatro PYMES, y una entidad sin ánimo de lucro que organizativamente correspondería también con una PYME. Todas ellas fueron propuestas por una empresa consultora especializada en proyectos de BI ubicada en el Parque Científico de la Universidad de Valladolid. A continuación se describen cada una de las organizaciones, manteniendo el anonimato por expreso deseo de sus máximos responsables:

[7] https://www.gartner.com/en/newsroom/press-releases/2018-12-06-gartner-data-shows-87-percent-of-organizations-have-low-bi-and-analytics-maturity
Fecha de consulta: 10 de junio de 2019

- Empresa A: Empresa Consultora de ámbito nacional dedicada a la búsqueda de financiación público privada para proyectos de inversión y de I+D+i. Cuenta con varias oficinas en la Comunidad

- Empresa B: Empresa Consultora con clientes en todo el país, que realiza la dirección de proyectos de sus clientes, operando como una PMO (Project Management Office) externa. Solamente un emplazamiento en Castilla y León.

- Empresa C: Entidad sin ánimo de lucro dedicada a mejorar la calidad de vida de personas con discapacidad y la de sus familias, así como la defensa de sus derechos. Cuenta con múltiples oficinas en Castilla y León., que es donde circunscribe su actividad.

- Empresa D: Empresa del sector cárnico con varias plantas, que actúa también en mercados exteriores además del mercado nacional.

- Empresa E: Empresa del sector lácteo, productora de quesos, que opera internacionalmente de manera muy importante (más del 50% de su producción) y con una sola planta en la Comunidad.

Las PYMES seleccionadas para el estudio de investigación pertenecen a diversos sectores de actividad y no son competidoras entre ellas. Todas ellas se encontraban saneadas y en crecimiento cuando se iniciaron los proyectos de implementación y a diferencia de otras organizaciones predigitales pertenecientes a industrias tradicionales, ninguna de las organizaciones objeto de estudio había experimentado presión alguna para iniciar un proceso de transformación digital.

Aunque en el momento de iniciar los proyectos las empresas no sienten ninguna amenaza inminente que les obligue al cambio, la operativa diaria de funcionamiento denotaba una problemática en el tratamiento de los datos y la información. En reuniones iniciales realizadas con los intervinientes en los proyectos se observó que casi todas utilizaban el programa microsoft excel de forma manual para la gestión de los datos, alguna utilizaba el programa microsoft Outlook para la gestión de tareas de equipos, pero no para hacer una planificación del trabajo a realizar. Estos programas informáticos permiten gestionar datos, pero cuando el volumen se hace demasiado grande ya no son gestores ni eficientes ni eficaces y empieza a ser necesario contar con otras herramientas como las que proporciona un sistema BI.

## 4.1. Empresa A

La empresa realizaba los informes para el Comité de Dirección a partir de hojas Excel e información que había que recopilar y procesar manualmente. Además, utilizaban Microsoft Outlook para la gestión de tareas de equipos, pero no para hacer una planificación del trabajo a realizar. El volumen de datos era difícilmente manejable y la información obtenida dejo de ser fiable.

El proyecto consistió en desarrollar una aplicación a medida para automatizar la recogida de información de todos los proyectos, y poder visualizar los datos mediante cuadros de mando creados ad hoc para la empresa. El proyecto se desarrolló durante 12 meses, entre 2018 y 2019.

El proyecto lo dirigió el director de desarrollo de proyectos de I+D+i y contó con la participación de los Responsables de Producción, Recursos Humanos y una persona de Administración. El CEO de la empresa estaba parcialmente comprometido con la implantación de las soluciones previstas y asistió a 2 reuniones con los consultores.

La empresa actualmente no cuenta con un ERP, pero dispone de un entorno colaborativo que permite compartir alguna información operativa. Durante el proyecto se realizó la racionalización de algunos de los procesos operativos. No hay un departamento TIC en la empresa.

## 4.2. Empresa B

La empresa contaba con un ERP que se utilizaba únicamente como programa contable y CRM básico para gestión de clientes. A partir de los datos introducidos obtenían unas métricas para medir el rendimiento a nivel departamental, aunque la obtención de los datos era difícil y existían dudas sobre su fiabili ad. El personal carecía de formación específica sobre herramientas de gestión de datos y su utilización.

El objetivo fue desarrollar una solución integrada para la Oficina Digital de la empresa: procesos de producción para la gestión de los proyectos y de gestión de clientes, administración y facturación con los CRM y ERP seleccionados. Todo ello totalmente integrado y conectado. La monitorización y el control de las actividades y el seguimiento de los proyectos para los clientes a través de cuadros de mando. El plazo establecido fue de 9 meses, desde octubre de 2018 a junio de 2019.

El proyecto lo dirigió el propio CEO de la empresa, cuyo compromiso fue total. En el proyecto participaron de forma activa y permanente, los responsables de producción y administración, que se encargaron de la implantación y desarrollo de los trabajos a realizar. Otros participantes en el proyecto fueron el Responsable de marketing y diseño y un Gestor de proyectos. Todos ellos muy involucrados en la implantación.

En el momento de finalizació  del proyecto la empresa cumplía algunos de los requisitos del cuarto nivel de madurez según el modelo de Gartner, aunque el CEO estaba planificando avanzar al siguiente estadio en la madurez digital de su empresa.

### 4.3. Empresa C

La organización tiene delegaciones en todas las provincias de Castilla y León. En Valladolid se centraliza la gestión de todas las actividades: cursos de formación, solicitud de ayudas a la administración, gestión de subvenciones, etc. y reciben múltiples archivos Excel de las delegaciones con información que han de utilizar para el desarrollo de todas las actividades previstas. La sede central de Valladolid se encontraba con la dificultad de homogeneizar todos los datos recibidos y sistematizar las tareas a realizar.

Se propuso la creación de un entorno de trabajo colaborativo (ECAM) compartido con las delegaciones de la organización. Se desarrollaron cuadros de mando y procesos ETL para integrar los datos suministrados de diferentes orígenes.

El interlocutor principal fue el Gerente, que además fue el promotor y estuvo algo comprometido para conseguir el éxito del proyecto. Entre enero y mayo de 2019 se programaron reuniones con el coordinador técnico, el Responsable de administración y dos técnicos de administración, de dos delegaciones territoriales diferentes, encargados de la confección y envío de los archivos Excel con la toma de datos. La duración del proyecto fue de 8 meses, de septiembre de 2018 a mayo de 2019.

### 4.4. Empresa D

A través de una aplicación informática móvil los encargados de las plantas recopilaban datos a tiempo real que se descargaban directamente en un archivo Excel, pero carecía de un sistema de BI que permitiera explotar la información que diariamente se almacenaba de todos los centros de producción. La información obtenida de los datos descargados en Excel no permitía la toma de decisiones eficaces, ni era fiable. La empresa contaba con un ERP para la gestión económica financiera.

El proyecto consistió en desarrollar unos cuadros de mando para monitorear los procesos productivos de las diferentes plantas, con actualización automática diaria a partir de los sistemas transaccionales existentes. Los objetivos específicos que se buscaban eran facilitar la información de los datos a través de cuadros de mando y desarrollar herramientas de gestión para la toma de decisiones ágiles y eficaces

El interlocutor principal del proyecto, con el que planificamos las reuniones, fue el Responsable de calidad e innovación. Además, se programaron reuniones con el CEO y dos empleados encargados de la toma de datos en dos centros de producción diferentes. El proyecto se ejecutó en dos meses, mayo y junio de 2019.

### 4.5. Empresa E

La empresa contaba con un sistema ERP para la gestión diaria de clientes y de proveedores, con el que se realizaba la facturación. Además, llevaban un control de la gestión de inventarios de repuestos e incidencias con el programa Microsoft excel. Carecía de estrategia de TD y de soluciones de BI.

El proyecto consistió en desarrollar los cuadros de mando para la gestión del mantenimiento de la empresa, además de un sistema automático de captación de los datos de la operativa diaria. El interlocutor principal del proyecto, con el que planificamos las reuniones fue el Director comercial. Además, se programaron reuniones con el Responsable de ventas y de administración y un empleado encargado del mantenimiento en la planta. La duración del proyecto fue de 4 meses, de enero a abril de 2019.

En el momento de finalización del proyecto la empresa se encontraba en el segundo nivel de madurez según el modelo de Gartner. Tenían previsto avanzar en su madurez digital generando cuadros de mando a partir de información obtenida del ERP ya instalado.

Una vez realizada la implantación de los proyectos de BI en las organizaciones surge la necesidad de conocer los resultados alcanzados y determinar las lecciones aprendidas del proceso de implementación; para ello se procedió a la observación de los resultados obtenidos, la recopilación de documentos del proyecto para su examen y valoración, y la identificación de hallazgos relevantes en la implementación de los proyectos de BI en PYMES.

**Tabla 3** Madurez digital de BI, situación inicial y final de las PYMES objeto de estudio.

| | Empresa A | | Empresa B | | Empresa C | | Empresa D | | Empresa E | |
|---|---|---|---|---|---|---|---|---|---|---|
| | Inicial | Final | Inicial | Final | Inicial | Final | Inicial | Final | Inicial | Final |
| **Nivel 1: Básico** | | | | | | | | | | |
| Datos desestructurados y mal interpretados | | | | | | | | | | |
| Datos en hojas de cálculo o excel e informes limitados | X | | | | X | | | | | |
| No medición de rendimiento y falta de conciencia importancia BI | X | | | | X | | | | | |
| Gestión en departamento TIC | | O | | | X | | | | | |
| **Nivel 2: Oportunista** | | | | | | | | | | |
| Métricas medir rendimiento por Departamento, no a nivel de empresa | | O | X | | | O | | | | |
| Difícil acceso a los datos y herramientas | X | | X | | X | | | | | |
| Aplicaciones de tratamiento de datos estándar | | O | X | O | | O | X | O | X | O |
| Usuarios no tienen formación especifica | X | | X | | X | | X | | | |
| No se confía en la información | X | | | | | | | | X | |
| **Nivel 3: Sistemático** | | | | | | | | | | |
| Existen expertos en BI y formación básica para usuarios | | O | | O | | O | X | | X | O |
| Responsables BI por unidad de negocio o departamento TIC | | | X | O | | | | | X | O |
| Indicadores por unidad de negocio o departamento TIC | | O | | O | | | X | | X | |
| No existe integración de datos a nivel de empresa | | | | | | | X | | X | |
| Se crea un BICC (Business Intelligence Competence Center) | | | | | | | | O | | O |
| **Nivel 4: Diferenciador** | | | | | | | | | | |
| Existe estrategia de desarrollo de BI a nivel de empresa | | | | | | | | | | |
| El BICC tiene suficientes recursos | | | | | | | | | | |
| Disponibilidad de información a todos los niveles | | | | O | | | | | | |
| Extensión del BI a proveedores y clientes | | | | | | | | O | | O |
| Se confía en la información y es soporte para la toma de decisiones | | | | | | | | O | | O |
| Usuarios formados en procesamiento de datos | | | | | | | | O | | O |
| **Nivel 5: Transformador** | | | | | | | | | | |
| Procesos BI en todas las unidades de negocio y en cultura de la empresa | | | | | | | | | | |
| BICC proactivo y dinámico | | | | | | | | | | |
| Información fiable y utilizada a todos los niveles de la empresa | | | | | | | | | | |
| Usuarios bien formados, utilizan la información para la creación de valor | | | | | | | | | | |
| BI disponible para socios, proveedores y clientes | | | | | | | | | | |

## 5. Discusión y resultados

Todas las empresas analizadas estaban desarrollando un proyecto BI, pero ninguna de ellas tenía una estrategia específica para BI (en el modelo de madurez de Gartner esto corresponde a un nivel 4). Este es un error habitual en las empresas: iniciar la realización de un Proyecto de BI sin contar con una estrategia de TD (Hess et al. 2016). De hecho, las empresas se centran en muchos casos en problemas específicos con tecnologías concretas (Kane et al. 2015). Aunque la estrategia de TD debería ser independiente de la "estrategia de tecnologías de información" (IT Strategy) en un buen número de empresas ambas se confunden o integran como una sola cosa. La estrategia de TD debería alinearse con las estrategias al nivel funcional y operacional para realmente optimizar el valor de las inversiones y los proyectos que la organización decida emprender.

En este sentido, contar con modelos de Madurez empresarial puede ayudar a definir la estrategia (Chanias et al. 2019) buscando lograr las cotas más altas para reinventar o redefinir los negocios y la empresa. En cualquier caso aún no existe una respuesta clara a la pregunta de cómo desarrollar una estrategia de TD.

### 5.1. RQ1: ¿Son los modelos de madurez en BI definidos po consultoras apropiados para aplicarlos a PYMES?

Sería muy difícil que una PYME pudiera alcanzar el nivel 3 del modelo de Madurez de Gartner, en el que además de contar con un departamento TIC, la empresa debería contar con un Business Intelligence Competence Center, y además expertos y responsables en BI en todos los departamentos. En concreto, de las empresas estudiadas, todas ellas con más de 5 trabajadores, ninguna cuenta con un departamento TIC. En todos los casos de estudio las empresas se encontraban en el nivel básico de madurez de Gartner en BI. Además, las empresas concluyeron que contar con una estrategia BI es clave para progresar y avanzar en el proceso de TD, por lo que esta condición debería situarse para superar el nivel 2, y no en el nivel 4. Esto implicaría renombrar este nivel denominado "Oportunista" para evidenciar la existencia de una planificación consciente

Otra característica de las PYMES es la existencia de estructuras poco jerarquizadas verticalmente, contando en general con estructuras muy planas, en las que los trabajadores desempeñan varios roles y la especialización no es muy elevada. Por ello la consideración del capital humano debería ser diferente en los modelos de madurez en BI y TD para las PYMES.

Los modelos de madurez digital más utilizados no se ajustan en ningún caso a la situación de las PYMES, que deberían tener un modelo más adecuado a sus características singulares para que puedan ser de utilidad en la definición de una hoja de ruta. El punto de partida debería ser el análisis de las competencias individuales y organizacionales deseables, y la definición de los niveles conforme al progreso en el desarrollo de esas competencias.

### 5.2. RQ2: ¿Son de utilidad estos modelos para analizar y seleccionar los proyectos de BI más adecuados y para evaluar su impacto al concluir su ejecución?

En los casos analizados ninguna de las entidades emprendió los proyectos de BI dentro de una estrategia clara y definida de TD. De hecho ninguna de ellas contaba con una estrategia de TD global de empresa. Por tanto, la mayoría iniciaron sus proyectos porque detectaron una necesidad operacional en sus organizaciones que querían solucionar y se decantaron por un proyecto de BI como apoyo para mejorar la toma de decisiones y el desempeño de la organización.

Nos basamos en ello para afirmar que, al menos inicialmente, los proyectos de implementación de BI en las organizaciones objeto de estudio encajan con episodios de TD conforme al modelo propuesto por Chanias et al. (2019). Se realizan actividades concretas de TD, que en nuestro caso fueron proyectos de BI, de acuerdo a las necesidades identificadas por las organizaciones (Burgelman et al. 2018), pero sin que exista una estrategia deliberada que englobe a toda la organización. En tanto que el único proyecto de TD era una implantación de proyecto de BI, podemos afirmar que las PYMES objeto de estudio tenían al menos inicialmente una estrategia abierta respecto a la transformación digital (Morton et al. 2017; Tavakoli et al. 2017).

En las reuniones realizadas con los entrevistados, la mayoría, indican que los presupuestos aprobados para la implantación de los proyectos son insuficientes para cubrir los objetivos necesarios para una implantación exitosa de los sistemas de BI. Uno de los entrevistados indica: "los directivos de la empresa consideran los gastos derivados del proyecto de BI como costes a corto plazo, que al no ser de carácter obligatorio pueden considerarse innecesarios, más aún cuando la empresa está actualmente en una situación de crecimiento de su cifra de negocio". Según Solis and Littleton (2017) cuando las empresas invierten en proyectos de transformación digital, es habitual que los gastos generados se consideren costes a corto plazo, con recursos y presupuestos muy limitados, en lugar de considerarse inversiones a largo plazo para la creación de valor.

Por un lado, un presupuesto insuficiente hace peligrar el éxito de la implantación de los proyectos de BI y por otro impide que se pueda desarrollar una estrategia de transformación digital planificada y alineada con la estrategia de la empresa. Además, la falta de obligatoriedad de la implementación del BI y que las PYMES no perciban como ob-jetivo prioritario la TD hace que los presupuestos dedicados no sean suficientes para cubrir con los objetivos deseados y en ocasiones sea necesario reducir el alcance del proyecto.

Un modelo de madurez que se desarrollara desde una óptica de desarrollo de competencias organizacionales e individuales sería más adecuado que los actuales modelos de madurez propuestos por las grandes consultoras, pensando en su aplicación a grandes corporaciones.

### 5.3. RQ3: ¿En qué medida los proyectos de BI contribuyen a alcanzar la madurez digital, concepto propuesto por Kane (2017) para referirse a la capacidad de responder al cambio de manera temprana?

Si bien la aceptación por parte de las empresas del BI es considerable y la importancia de los sistemas de BI es ampliamente reconocida, hay pocos estudios que hayan investigado los factores críticos que afectan el éxito de la implementación de proyectos de BI (Yeoh and Koronios 2010). Ruiz et al. (2018) justifica que los proyectos de BI contribuyen a elevar la resiliencia organizacional, lo que está alineado con el concepto de madurez digital. En todos los casos analizados los máximos responsables experimentaron un cambio importante, materializado en el interés por desarrollar una Estrategia de Transformación Digital en la que el BI tenga un papel central para contribuir a una toma de decisiones más informada.

Aunque hemos documentado la formulación e implementación de unos proyectos de BI, no existen garantías de que las organizaciones sigan avanzando en su TD futura, puesto que la TD siempre está en proceso y el futuro es incierto. Faltan estudios de investigación del proceso, el éxito, fracasos y los riesgos que han de asumir las PYMES en su proceso de TD. Aún no existe una forma de evaluar el avance en la TD, ni cuando la TD de una PYME puede considerarse un éxito.

El resultado de los proyectos de implementación de BI en los casos de estudio son estrategias de transformación digital realizada (Mintzberg and Waters 1985), que de acuerdo al modelo propuesto por Chanias et al. (2019) representa una instantánea mental, en un momento concreto, de lo que ha logrado la organización en términos de elaboración de su estrategia de TD. Una estrategia de transformación digital realizada también sirve de entrada para futuras actividades de estrategia, formando parte de un ciclo de retroalimentación continuo. Sería conveniente avanzar en la investigación con las PYMES objeto de estudio para demostrar que, según el modelo propuesto, deben aprender a elaborar de forma iterativa los fundamentos de la estrategia de TD. También, solicitamos más estudios que investiguen si las estrategias de transformación digital realizadas representan estructuras socialmente logradas que se reproducen a través de la acción social (Giddens 1984), desarrollando la estrategia digital a través de distintas prácticas de estrategia de TD.

Por otro lado, aunque el campo de la aceptación de las tecnologías digitales está bien investigado, no sucede igual con el uso posterior a la adopción de las tecnologías digitales (Jasperson et al. 2005). La aceptación inicial es importante para el éxito en la implementación de los proyectos de BI (Delone and McLean 1992) aunque no es condición suficiente para un uso posterior integrado con los procesos, rutinas y la estrategia empresarial (Shanks et al. 2012). Existe la posibilidad de que el uso disminuya con el tiempo o sea un uso desconectado. Por otro lado, Grublješič and Jaklič (2015b) indican que en ninguna de las organizaciones que fueron estudiadas la alta dirección utilizó el BI como se esperaba: la utilización fue inferior al objetivo deseado. La investigación sobre la aceptación de la tecnología y el uso posterior a la implementación de los proyectos de BI, así como el uso que la alta dirección hace del BI, podrá ayudar a determinar el éxito de la implementación de soluciones de BI en PYMES.

## 6. Conclusiones

Los resultados obtenidos pueden ser útiles para PYMES y organizaciones que estén en el nivel básico de digitalización, y que quieran implementar proyectos concretos de TD o iniciar una estrategia de TD. Los modelos de madurez de BI actuales no son realmente de utilidad, y sería preciso realizar una aproximación diferente mediante un enfoque basado en competencias individuales y organizacionales.

Las PYMES que abordan un proyecto de BI lo hacen inicialmente para resolver una necesidad operacional, aunque a la conclusión del proyecto descubren la importancia y necesidad de contar con una estrategia de TD en general, y de BI en particular. Contar con modelos de madurez que ayuden a establecer el plan de brechas para las organizaciones será de mucha utilidad para el análisis y selección de los proyectos de BI prioritarios.

Los proyectos de BI contribuyen a la madurez digital de la organización, aunque no sean diseñados y aprobados dentro de una estrategia bien definida, aunque se realicen con un presupuesto inferior al que sería necesario asignar, y aunque finalmente no se explote el 100% de su potencial

## 7. Agradecimientos

Este trabajo ha sido realizado gracias a la financiación mediante un Contrato de Investigación al amparo del Art. 83 financiado por la empresa CGPROTEC: "Proyectos de Modernización Empresarial para la Industria 4.0 y el Desarrollo de la Empresa Digital: Business Intelligence Consultoría".

## 8. Bibliografía

Bharadwaj, A., OA El Sawy, PA Pavlou, and N. Venkatraman. 2013. "Digital Business Strategy: Toward a next Generation of Insights." MIS Quarterly 471–82.

Burgelman, Robert A., Steven W. Floyd, Tomi Laamanen, Saku Mantere, Eero Vaara, and Richard Whittington. 2018. "Strategy Processes and Practices: Dialogues and Intersections." Strategic Management Journal 39(3):531–58.

Caserio, Carlo and Sara Trucco. 2018. Enterprise Resource Planning and Business Intelligence Systems for Information Quality: An Empirical Analysis in the Italian Setting. Vol. 21.

Chanias, Simon, Michael D. Myers, and Thomas Hess. 2019. "Digital Transformation Strategy Making in Pre-Digital Organizations: The Case of a Financial Services Provider." Journal of Strategic Information Systems 28(1):17–33.

Creswell, John. 2013. "Qualitative Inquiry & Research Design: Choosing Among Five Approaches." SAGE Publications 11.

Dedić, Nedim and Clare Stanier. 2016. Measuring the Success of Changes to Existing Business Intelligence Solutions to Improve Business Intelligence Reporting. Vol. 268.

Delone, William and Ephraim McLean. 1992. "Information Systems Success: The Quest for the Dependent Variable." Information Systems Research 3:60–95.

Elbashir, Mohamed Z., Philip A. Collier, and Michael J. Davern. 2008. "Measuring the Effects of Business Intelligence Systems: The Relationship between Business Process and Organizational Performance." International Journal of Accounting Information Systems 9(3):135–53.

Fitzgerald, M., N. Kruschwitz, D. Bonnet, and M. Welch. 2014. "Embracing Digital Technology: A New Strategic Imperative." MIT Sloan Management Review 55.

Giddens, A. 1984. The Constitution of Society: Outline of the Theory of Structuration. University of California Press.

Glazer, R. 1993. "Measuring the Value of Information: The Information-Intensive Organization." IBM Systems Journal 32(1):99–110.

Grublješič, Tanja and Jurij Jaklič. 2015a. "Business Intelligence Acceptance: The Prominence of Organizational Factors." Information Systems Management 32(4):299–315.

Grublješič, Tanja and Jurij Jaklič. 2015b. "Conceptualization of the Business Intelligence Extended Use Model." Journal of Computer Information Systems 55(3).

Hannula, Mika and Virpi Pirttimaki. 2003. "Business Intelligence Empirical Study on the Top 50 Finnish Companies." Journal of American Academy of Business.

Hess, Thomas, Christian Matt, Alexander Benlian, and Florian Wiesböck. 2016. "Options for Formulating a Digital Transformation Strategy." MIS Quarterly Executive 40(2):16–24.

Jasperson, Jon, Pamela E. Carter, and Robert W. Zmud. 2005. "A Comprehensive Conceptualization of Post-Adoptive Behaviors Associated with Information Technology Enabled Work Systems." MIS Quarterly: Management Information Systems.

Kane, G. C. 2017. "Digital Maturity, Not Digital Transformation." MIT Sloan Management Review.

Kane, G. C., Doug Palmer, A. N. Philips, David Kiron, and Natasha Buckley. 2015. "Strategy, Not Technology, Drives Digital Transformation." MIT Sloan Management Review and Deloitte University Press (57181):27.

Klein, Heinz K. and Michael D. Myers. 1999. "A Set of Principles for Conducting and Evaluating Interpretive Field Studies in Information Systems." MIS Quarterly: Management Information Systems 23(1):67–94.

Legner, Christine, Torsten Eymann, Thomas Hess, Christian Matt, Tilo Böhmann, Paul Drews, Alexander Mädche, Nils Urbach, and Frederik Ahlemann. 2017. "Digitalization: Opportunity and Challenge for the Business and Information Systems Engineering Community." Business and Information Systems Engineering 59(4):301–8.

Li, Xixi, J. J. Po-An Hsieh, and Arun Rai. 2013. "Motivational Differenc s Across Post-Acceptance Information System Usage Behaviors: An Investigation in the Business Intelligence Systems Context." Information Systems Research (May 2015).

Lorenzo Ochoa, Oswaldo. 2016. "Modelos De Madurez Digital: ¿En Qué Consisten Y Qué Podemos Aprender De Ellos?" Boletin De Estudios Economicos 71(219):573–90.

Mintzberg, Henry and James A. Waters. 1985. “Of Strategies, Deliberate and Emergent.” Strategic Management Journal 6(3):257–72.

Morton, Josh, Alex Wilson, Robert Galliers, and Marco Marabelli. 2017. “Open Strategy and IT : A Review and Research Agenda.” 33rd EGOS Colloquium (July):0–24.

Myers, MD. 2013. “Qualitative Research in Business and Management.”

Myers, Michael and Michael Newman. 2007. “The Qualitative Interview in IS Research: Examining the Craft.” Information and Organization 17:2–26.

P. Luhn, H. 1958. “A Business Intelligence System.” IBM Journal October:314–19.

Papachristodoulou, Ekavi, Margarita Koutsaki, and Efstathios Kirkos. 2017. “Business Intelligence and SMEs: Bridging the Gap.” Journal of Intelligence Studies in Business 7(1):70–78.

Peppard, Joe, Robert D. Galliers, and Alan Thorogood. 2014. “Information Systems Strategy as Practice: Micro Strategy and Strategizing for IS.” The Journal of Strategic Information Systems 23(1):1–10.

Pettigrew, Andrew M. 1987. “Context and Action in the Transformation of the Firm.” Journal of Management Studies.

Pettigrew, Andrew M. 2012. “Context and Action in the Transformation of the Firm: A Reprise.” Journal of Management Studies 49(7):1304–28.

Piccinini, E., R. W. Gregory, and L. M. Kolbe. 2015. “Changes in the Producer-Consumer Relationship-Towards Digital Transformation.” Association for Information Systems AIS Electronic Library (AISeL) 109.

Ponelis, Shana R. 2015. “Using Interpretive Qualitative Case Studies for Exploratory Research in Doctoral Studies: A Case of Information Systems Research in Small and Medium Enterprises.” International Journal of Doctoral Studies 10:535–50.

Popovič, Aleš, Ray Hackney, Pedro Simões Coelho, and Jurij Jaklič. 2012. “Towards Business Intelligence Systems Success: Effects of Maturity and Culture on Analytical Decision Making.” Journal of Decision Support Systems.

Reinschmidt, Joerg and Allison Francoise. 2000. “Business Intelligence Certification Guide.” IBM International Technical Support Organization.

Ruiz-Martín, C, Villafáñez, F; López-Paredes, A., Wainer, G (2017). "Impact Of Business Intelligence In Organizational Resilience". INFORMS Annual Meeting. Houston.

Sánchez-Marín, Gregorio, Antonio Sánchez, Rosa María Batista-Canino, Antonio Calvo-Flores Segura, Ignacio Danvila del Valle, Julio Diéguez-Soto, Escobar Bernabé, Domingo Garcia-Perez-de-Lema, José María Gómez Gras, Ricardo Hernández Mogollón, Juan Jimenez-Moreno, Francisco Javier Martínez García, Pablo Antonio Muñoz Gallego, C. Lázaro, and Francisco Somohano. 2018. Análisis Estratégico Para El Desarrollo de La PYME En España. Informe PYME España 2018. FAEDPYME.

Schultze, Ulrike. 2000. “A Confessional Account of An Ethnography about Knowledge Work.” MIS Quarterly 24:3–41.

Shanks, G., N. Bekmamedova, Frédéric Adam, and Mary Daly. 2012. “Embedding Business Intelligence Systems within Organisations.” Journal of Frontiers in Artificial Intelligence and Applications 238:113–24.

Solis, Brian and Aubrey Littleton. 2017. “The 2017 State of Digital Transformation.” Research Report. Altimeter, a Prophet Company, 1–38.

Stone, Merlin David and Neil David Woodcock. 2014. “Interactive, Direct and Digital Marketing: A Future That Depends on Better Use of Business Intelligence.” Journal of Research in Interactive Marketing 8(1):4–17.

Tavakoli, Asin, Daniel Schlagwein, and Detlef Schoder. 2017. “Open Strategy: Literature Review, Re-Analysis of Cases and Conceptualisation as a Practice.” The Journal of Strategic Information Systems 26:163–84.

Trkman, Peter, Kevin McCormack, Marcos Paulo Valadares De Oliveira, and Marcelo Bronzo Ladeira. 2010. “The Impact of Business Analytics on Supply Chain Performance.” Decision Support Systems 49(3):318–27.

Tvrdíková, Milena. 2007. “Support of Decision Making by Business Intelligence Tools.” Pp. 364–68 in 6th International Conference on Computer Information Systems and Industrial Management Applications, CISIM.

Venkatesh, V., MG Morris, GB Davis, and FD Davis. 2003. “User Acceptance of Information Technology: Toward a Unified View.” MIS Quarterly.

Venkatesh, Viswanath and Hillol Bala. 2008. “Technology Acceptance Model 3 and a Research Agenda on Interventions.” Decision Sciences 39(2):273–315.

Walsham, Geoff. 1993. "Interpreting Information Systems in Organizations." The Journal of the Operational Research Society 44.

Walsham, Geoff. 1995. "Interpretive Case Studies in IS Research: Nature and Method." European Journal of Information Systems 4(2):74–81.

Westerman, G., C. Calméjane, D. Bonnet, P. Ferraris, and A. McAfee. 2011. "Digital Transformation: A Roadmap for Billion-Dollar Organizations." MIT Center for Digital Business and Capgemini Consulting 1:1–68.

Yeoh, William and Andy Koronios. 2010. "Critical Success Factors for Business Intelligence Systems." Journal of Computer Information Systems 50(3):23–32.

Zeng, Li, L. Xu, Z. Shi, M. Wang, and W. Wu. 2006. "Techniques, Process, and Enterprise Solutions of Business Intelligence." Pp. 4722–26 in 2006 IEEE International Conference on Systems, Man and Cybernetics. Vol. 6.